\documentclass[12pt]{article}
 \usepackage{graphicx}
 \usepackage[cp1251]{inputenc}

\begin{document}
 \noindent {\footnotesize\it
   Astronomy Letters, 2022, Vol. 48, No. 3, pp. 169–177}
 \newcommand{\dif}{\textrm{d}}

 \noindent
 \begin{tabular}{llllllllllllllllllllllllllllllllllllllllllllll}
 & & & & & & & & & & & & & & & & & & & & & & & & & & & & & & & & & & & & & &\\\hline\hline
 \end{tabular}

  \vskip 0.5cm
  \bigskip
\centerline{\bf\Large Estimation of Galactic Spiral Density  }
\centerline{\bf\Large Wave Parameters Based on the Velocities }
 \centerline{\bf\Large of OB2 stars from the Gaia EDR3 Catalogue}
 \bigskip
 \bigskip
 \centerline{\bf
            V. V. Bobylev\footnote[1]{e-mail: vbobylev@gaoran.ru} (1) and
            A. T. Bajkova (1),
            }
 \bigskip
 \centerline{\small\it(1)
 Pulkovo Astronomical Observatory of Russian Academy of Sciences, St. Petersburg, Russia}
 \centerline{\small Received November 26, 2021; revised December 5, 2021; accepted December 28, 2021}

 \bigskip
 \bigskip
{\bf Abstract}---We have analyzed the kinematics of 9750 OB2 stars with proper motions and parallaxes
selected by Xu et al. from the Gaia EDR3 catalogue. The relative parallax errors for these stars do not
exceed 10\%. Based on the entire sample of stars, we have found the velocities $(U,V)_\odot=(7.17,7.37)\pm(0.16,0.24)$ km s$^{-1}$ and the components of the angular velocity of Galactic rotation: $\Omega_0 =29.700\pm0.076$~km s$^{-1}$ kpc$^{-1}$, $\Omega^{'}_0 =-4.008\pm0.022$~km s$^{-1}$ kpc$^{-2}$, and $\Omega^{''}_0 =~0.671\pm0.011$~km s$^{-1}$ kpc$^{-3}$, where
the linear rotation velocity of the Galaxy at the solar distance is $V_0=240.6\pm3.0$~km s$^{-1}$ for the adopted $R_0=8.1\pm0.1$~kpc. There are 1812 OB2 stars with measured line-of-sight velocities, and the space velocities $V_R$ and $\Delta V_{circ}$ have been calculated from them. Based on a spectral analysis independently for the radial and residual tangential velocities, we have obtained the following estimates: $f_R=4.8\pm0.7$~km s$^{-1}$, $f_\theta=4.1\pm0.9$~km s$^{-1}$, $\lambda_R=2.1\pm0.2$~kpc, $\lambda_\theta=2.2\pm0.4$~kpc, $(\chi_\odot)_R=-116\pm12^\circ$, and $(\chi_\odot)_\theta=-156\pm14^\circ$ for the adopted four-armed ($m = 4$) spiral pattern. Thus, both velocity perturbation amplitudes are nonzero at a high significance level.

\bigskip\noindent
{\bf DOI:} 10.1134/S1063773722020025


\newpage
\section*{INTRODUCTION}
Stars of spectral types O and early B are very young (several Myr) massive (more than 10$M_\odot$)
high-luminosity stars. Owing to these properties, they are of great importance for studying the structure and kinematics of the Galaxy on various scales.

In this paper our main interest is to estimate the parameters of the Galactic spiral density wave. We
estimate these parameters within the linear theory of spiral structure by Lin and Shu (1964). Various
samples of OB stars, HII regions (where stars of spectral type O are the central exciting ones), and
OB associations have repeatedly served to solve this problem (Cr\'ez\'e and Mennessier 1973; Byl and Ovenden 1978; Mel’nik et al. 2001; Fern\'andez et al. 2001; Zabolotskikh et al. 2002; Russeil 2003; Bobylev and Bajkova 2018).

Of course, there are also other tracers of the spiral structure in the Galaxy. These include, for example, long-period Cepheids, young open star clusters (OSCs), or maser sources. Such young objects are also commonly used to determine the structural and kinematic parameters of the Galactic spiral density
wave (Burton 1971; Mishurov et al. 1997; Mishurov and Zenina 1999; L\'epine et al. 2001; Popova and
Loktin 2005; Siebert et al. 2012; Griv et al. 2014; Griv and Jiang 2015; Rastorguev et al. 2017; Reid
et al. 2019; Loktin and Popova 2019; Xu et al. 2018, 2021; Barros et al. 2021).

The results of the kinematic analysis of OB stars depend strongly on the observational data quality.
The accuracies of the proper motions of stars improve continuously and have been measured at present for a large number of OB stars. The line-of-sight velocities for OB stars have been measured for a much
smaller number of them. The photometric distances of OB stars have been commonly used to analyze their
spatial distribution and kinematics. The situation has changed quite recently with the publication of more accurate and more reliable trigonometric parallaxes for millions of stars measured in the Gaia space experiment (Prusti et al. 2016).

At present, the Gaia EDR3 (Gaia Early Data Release 3, Brown et al. 2021) version of the catalogue
has been published, where, in comparison with the previous Gaia DR2 version (Brown et al. 2018), the
trigonometric parallaxes and proper motions were improved approximately by 30\% for $\sim$1.5 billion stars. The trigonometric parallaxes for $\sim$500 million stars were measured with errors less than 0.2 milliarcseconds (mas). For stars with magnitudes $G<15^m$ the random measurement errors of the proper motions lie in the range 0.02–0.04 mas yr$^{-1}$, and they increase dramatically for fainter stars. On the whole, the proper motions for about a half of the stars in the catalogue were measured with a relative error less than 10\%. There are no new line-of-sight velocity measurements in the Gaia EDR3 catalogue.

Xu et al. (2021) produced a large sample of OB2 stars with proper motions and trigonometric
parallaxes from the Gaia EDR3 catalogue. The goal of this paper is to redetermine the Galactic spiral
density wave parameters using the latest data on stars of spectral types O and B.

 \section*{METHOD}\label{method}
We have three stellar velocity components from observations: the line-of-sight velocity $V_r$ and the
two tangential velocity components $V_l=4.74r\mu_l\cos b$ and $V_b=4.74r\mu_b$ along the Galactic longitude $l$ and latitude $b$, respectively. All three velocities are expressed in km s$^{-1}$. Here, 4.74 is the dimension coefficient and $r$ is the stellar heliocentric distance in kpc. The proper motion components $\mu_l\cos b$ and $\mu_b$ are expressed in mas yr$^{-1}$. The velocities $U,V,W$ directed along the rectangular Galactic coordinate axes are calculated via the components $V_r, V_l, V_b$: 
 \begin{equation}
 \begin{array}{lll}
 U=V_r\cos l\cos b-V_l\sin l-V_b\cos l\sin b,\\
 V=V_r\sin l\cos b+V_l\cos l-V_b\sin l\sin b,\\
 W=V_r\sin b                +V_b\cos b,
 \label{UVW}
 \end{array}
 \end{equation}
where the velocity $U$ is directed from the Sun toward the Galactic center, $V$ is in the direction of Galactic rotation, and $W$ is directed to the north Galactic pole. We can find two velocities, $V_R$ directed radially away from the Galactic center and $V_{circ}$ orthogonal to it pointing in the direction of Galactic rotation, based on the following relations:
 \begin{equation}
 \begin{array}{lll}
  V_{circ}= U\sin \theta+(V_0+V)\cos \theta, \\
       V_R=-U\cos \theta+(V_0+V)\sin \theta,
 \label{VRVT}
 \end{array}
 \end{equation}
where the position angle $\theta$ obeys the relation $\tan\theta=y/(R_0-x)$, $x,y,z$ are the rectangular heliocentric coordinates of the star (the velocities $U,V,W$ are directed along the corresponding $x,y,z$ axes), and $V_0$ is the linear rotation velocity of the Galaxy at the solar distance $R_0$.

To determine the parameters of the Galactic rotation curve, we use one conditional equation with
the velocity component $V_l$ on the left-hand side. This equation was derived from Bottlinger’s formulas, where the angular velocity of Galactic rotation $\Omega$ is expanded into a series to terms of the second order of smallness in $r/R_0$:
 \begin{equation}\begin{array}{lll}
 V_l= U_\odot\sin l-V_\odot\cos l
 -r\Omega_0\cos b\\
 +(R-R_0)(R_0\cos l-r\cos b)\Omega^\prime_0\\
 +0.5(R-R_0)^2(R_0\cos l-r\cos b)\Omega^{\prime\prime}_0,\\
 +\tilde{v}_R \sin(l+\theta)+\tilde{v}_\theta \cos(l+\theta),
 \label{EQ-2}
 \end{array}\end{equation}
where $R$ is the distance from the star to the Galactic rotation axis, $R^2=r^2\cos^2 b-2R_0 r\cos b\cos l+R^2_0$. The velocities $(U,V,W)_\odot$ are the mean group velocity of the sample, are taken with the opposite sign, and reflect the peculiar motion of the Sun. Since the velocity $W_\odot$ cannot be determined well only from the components $V_l$, we take its value to be 7 km s$^{-1}$; $\Omega_0$ is the angular velocity of Galactic rotation at the solar distance $R_0$, the parameters $\Omega^{\prime}_0$ and $\Omega^{\prime\prime}_0$ are the corresponding derivatives of the angular velocity, and $V_0=R_0\Omega_0$.
 
In this paper $R_0$ is taken to be $8.1\pm0.1$~kpc, according to the review by Bobylev and Bajkova (2021), where it was derived as a weighted mean of a large number of present-day individual estimates. Note also the most accurate present-day individual measurement of R0 obtained by Abuter et al. (2019)
by analyzing a 16-year series of observations of the motion of the star $S2$ around the massive black
hole $Sgr A^*$ at the Galactic center, $R_0=8.178\pm0.013$\,(stat)$\pm0.022$\,(syst)~kpc.

We took into account the influence of the Galactic spiral density wave in Eq. (\ref{EQ-2}) based on the linear theory, in which the potential perturbation is in the form of a traveling wave (Lin and Shu 1964; Lin et al. 1969). This influence is taken into account in the form suggested by Cr\'ez\'e and Mennessier (1973). The influence of the spiral density wave in the radial velocities $V_R$ and residual tangential velocities $\Delta V_{circ}$ is determined by the velocities ?$\tilde{v}_R$ and $\tilde{v}_\theta$, respectively:
 \begin{equation}\begin{array}{lll}
      \tilde{v}_R =-f_R \cos \chi,\\
 \tilde{v}_\theta = f_\theta \sin \chi,
 \label{DelVRot}
 \end{array}\end{equation}
where
 $$ 
 \chi=m[\cot(i)\ln(R/R_0)-\theta]+\chi_\odot
 \label{chi}
 $$ 
is the phase of the spiral density wave, $m$ is the number of spiral arms, $i$ is the pitch angle of the spiral pattern ($i<0$ for a winding spiral), and $\chi_\odot$ is the Sun’s radial phase in the spiral density wave; $f_R$ and $f_\theta$ are the amplitudes of the radial and tangential velocity perturbations, which are assumed to be positive.
 
The minus sign before $\cos \chi$ in the first equation (\ref{DelVRot}) shows that at the center of the spiral arm in the region inside the solar circle $f_R$ is directed toward the Galactic center. This corresponds to the well-known illustration in Rohlfs (1977), where the directions of the velocity perturbations in the spiral density wave in the region inside the solar circle are specified.

The sign of the Sun’s phase angle $\chi_\odot$ can be both positive and negative, depending on the reference point. When measuring this angle from the Carina–Sagittarius spiral arm ($R\sim7$~kpc), it will be negative. When measured from the Perseus arm ($R\sim9.5$~kpc), the Sun’s phase angle will be positive.

The wavelength $\lambda$ (the distance between adjacent spiral arm segments measured along the radial direction) is calculated from the relation
\begin{equation}
 2\pi R_0/\lambda=m\cot(|i|).
 \label{a-04}
\end{equation}
Solving the system of conditional equations (\ref{EQ-2}) by the least-squares method (LSM), we can determine the velocities $(U,V)_\odot$, $\Omega_0$, $\Omega^{\prime}_0$, $\Omega^{\prime\prime}_0$, $\tilde{v}_R$, and $\tilde{v}_\theta$. To estimate the amplitudes of the velocity perturbations $f_R$ and $f_\theta$, we need to know the pitch angle $i$ and the Sun’s phase $\chi_\odot$ (or, according to Eq. (\ref{a-04}), the wavelength $\lambda$ and $\chi_\odot$).

This approach was implemented, for example, by Mishurov and Zenina (1999) while analyzing
Cepheids. To determine $f_R$ and $f_\theta$ by minimizing the $\chi^2$, two variables, $i$ and $\chi_\odot$, were varied in a fairly wide range. The values of $f_R$ and $f_\theta$ were searched for by Mel’nik et al. (2001) while studying the kinematics of OB associations and by Popova and Loktin (2005) while analyzing OSCs and OB stars in approximately the same way.

There is also another approach to determine such parameters of the spiral density wave as $f_R$, $f_\theta$, $\lambda$~(or $i$), and $\chi_\odot$ by analyzing the radial ($V_R$) and residual tangential ($\Delta V_{circ}$) velocities of stars. This approach is based on a modified spectral analysis (Bajkova and Bobylev 2012). Here, the residual velocities $\Delta V_{circ}$ were derived from the tangential velocities $V_{circ}$ by subtracting the rotation curve with predetermined parameters.

Let there be a series of measured velocities $V_{R_n}$ (these can be both radial ($V_R$) and tangential ($\Delta V_{circ}$) velocities), $n=1,\dots,N$, where $N$ is the number of objects. The objective of the spectral analysis is to extract a periodicity from the data series in accordance with the adopted model describing a spiral density wave with parameters $f,$ $\lambda$~(or $i$), and $\chi_\odot$.

Having taken into account the logarithmic behavior of the spiral density wave and the position angles
of the objects $\theta_n$, our spectral (periodogram) analysis of the series of velocity perturbations is reduced to calculating the square of the amplitude (power spectrum) of the standard Fourier transform (Bajkova and Bobylev 2012):
\begin{equation}
 \bar{V}_{\lambda_k} = \frac{1} {N}\sum_{n=1}^{N} V^{'}_n(R^{'}_n)
 \exp\biggl(-j\frac {2\pi R^{'}_n}{\lambda_k}\biggr),
 \label{29}
\end{equation}
where $\bar{V}_{\lambda_k}$ is the $k$th harmonic of the Fourier transform
with wavelength $\lambda_k=D/k$, $D$ is the period of the series being analyzed,
 \begin{equation}
 \begin{array}{lll}
 R^{'}_{n}=R_0\ln(R_n/R_0),\\
 V^{'}_n(R^{'}_n)=V_n(R^{'}_n)\times\exp(jm\theta_n).
 \label{21}
 \end{array}
\end{equation}
The sought-for wavelength $\lambda$ corresponds to the peak value of the power spectrum $S_{peak}$. The pitch angle of the spiral density wave is found from Eq. (\ref{a-04}). We determine the perturbation amplitude and phase by fitting the harmonic with the wavelength found to the observational data. The following relation can be used to estimate the perturbation amplitude:
 \begin{equation}
 f_R (f_\theta)=2\times\sqrt{S_{peak}}.
 \label{Speak}
 \end{equation}

In this paper we use both described methods. For this purpose, we apply the following approach. In the
first step, we seek the LSM solution of the system of conditional equations (\ref{EQ-2}) to estimate the six parameters $(U,V)_\odot,$ $\Omega_0,$ $\Omega^{\prime}_0$, and $\Omega^{\prime\prime}_0$. With these velocities we form the radial, $V_R$, and residual rotation, $\Delta V_{circ}$, velocities.

In the second step, we perform a spectral analysis of the velocities $V_R$ and $\Delta V_{circ}$ and find $f_R$, $f_\theta$, $\lambda$ and $\chi_\odot$. This method takes into account both the logarithmic behavior of the Galactic spiral structure and the position angles of the objects, which allows the velocities of the objects distributed in a wide range of Galactocentric distances to be analyzed accurately. In addition, by this method we can obtain estimates both only from the radial velocities, $\lambda_R$ and $(\chi_\odot)_R$, and from the residual tangential velocities, $\lambda_\theta$ and $(\chi_\odot)_\theta$.

In the third step, we seek the LSM solution of the system of conditional equations (\ref{EQ-2}) to estimate the
seven parameters $(U,V)_\odot,$ $\Omega_0,$ $\Omega^{\prime}_0$, $\Omega^{\prime\prime}_0$, $\tilde{v}_R$, and $\tilde{v}_\theta$. Given $\lambda$ and $\chi_\odot$ found in the second step, we obtain new estimates of $f_R$ and $f_\theta$.

 \section*{DATA}
We use the sample of OB stars for which the proper motions and trigonometric parallaxes were
taken by Xu et al. (2021) from the Gaia EDR3 catalogue. For this purpose, these authors identified
9750 stars of spectral types from O to B2 spectroscopically confirmed by Skiff (2014) with
the Gaia EDR3 catalogue. Stars with relative trigonometric parallax errors less than 10\% were
selected. No stars located higher than 300 pc above the Galactic plane were included in the sample. 
 
Xu et al. (2018) produced a sample of 5772 OB2 stars with kinematic parameters from the Gaia DR2
catalogue. Stars with relative trigonometric parallax errors less than 10\% were selected. For more than 2500 OB stars these authors took the line-of-sight velocities from the SIMBAD electronic database \footnote{http://simbad.u-strasbg.fr/simbad/}.

We identified the samples of OB stars from Xu et al. (2018) and Xu et al. (2021) and detected
1812 stars with line-of-sight velocities in the new sample. The line-of-sight velocities of the OB stars in the catalogue by Xu et al. (2018) are given relative to the local standard of rest and, therefore, we convert them back to the heliocentric ones with the known standard solar motion parameters $(U,V,W)_\odot=(10.3,15.3,7.7)$ km s$^{-1}$.

\begin{figure}[t]
{ \begin{center}
  \includegraphics[width=0.75\textwidth]{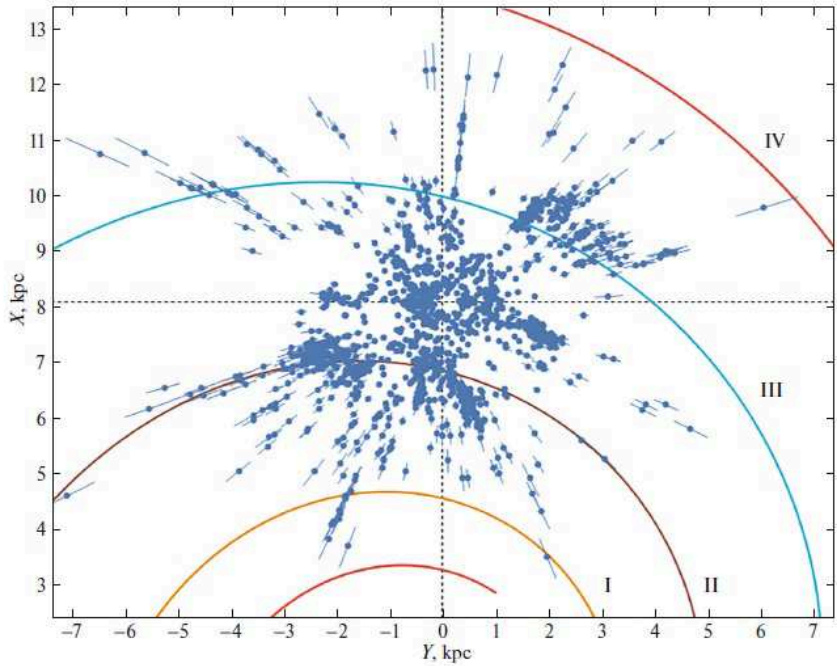}
  \caption{
Distribution of OB2 stars with line-of-sight velocities on the Galactic $XY$ plane; the four-armed spiral pattern with a pitch angle $i=-13^\circ$ is shown.
  }
 \label{f-1812-XY}
\end{center}}
\end{figure}

Figure~\ref{f-1812-XY} presents the distribution of 1812 OB2 stars in projection onto the Galactic $XY$ plane. We use the coordinate system in which the $X$ axis is directed from the Galactic center to the Sun and the direction of the $Y$ axis coincides with the direction of Galactic rotation. The four-armed spiral pattern with a pitch angle $i=-13^\circ$ (Bobylev and Bajkova 2014a) constructed with $R_0=8.1$~kpc is shown; the Roman numerals number the following spiral arm segments: Scutum (I) , Carina–Sagittarius (II), Perseus (III), and the Outer Arm (IV).

\begin{figure}[t]
{ \begin{center}
  \includegraphics[width=0.85\textwidth]{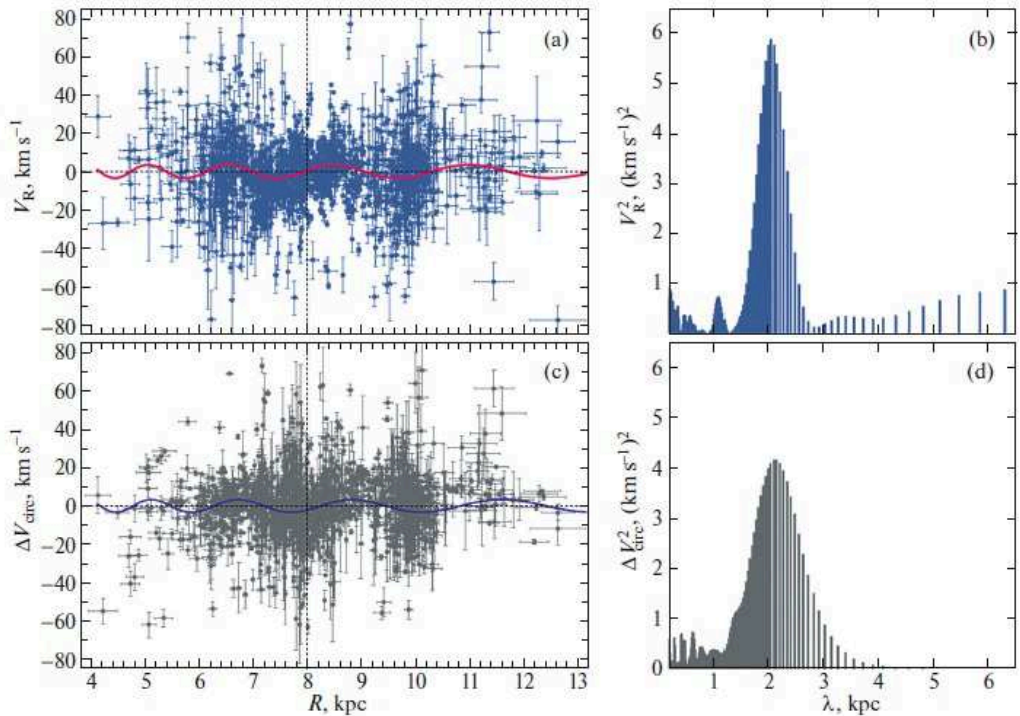}
  \caption{
(a) Radial velocities $V_R$ versus distance $R$ for the sample of OB2 stars with line-of-sight velocities, (b) the power spectrum of this sample, (c) the residual rotation velocities $\Delta V_{circ}$ of the stars from this sample, and (d) their power spectrum.
 }
 \label{f-1725-RT}
\end{center}}
\end{figure}

 \section*{RESULTS}
In the first step, the system of conditional equations (\ref{EQ-2}) is solved for five unknowns by the least squares method with weights of the form $w_l=S_0/\sqrt {S_0^2+\sigma^2_{V_l}}$, where $S_0$ is the “cosmic” dispersion and $\sigma_{V_l}$ is the dispersion of the observed velocities. $S_0$ is comparable to the root-mean-square residual $\sigma_0$ (the error per unit weight) in solving the conditional equations (\ref{EQ-2}). We adopted $S_0=10$~km s$^{-1}$. The system of equations was solved in several iterations using the $3\sigma$ criterion to eliminate the OSCs with large residuals. 
 
Based on the entire sample of 9750 OB2 stars with proper motions and parallaxes, we found the two velocities $(U,V)_\odot=(7.17,7.37)\pm(0.16,0.24)$~km s$^{-1}$ and the following components of the angular velocity of Galactic rotation:
 \begin{equation}
 \label{sol-I}
 \begin{array}{lll}
      \Omega_0 =29.700\pm0.076~\hbox{km s$^{-1}$ kpc$^{-1}$},\\
  \Omega^{'}_0 =-4.008\pm0.022~\hbox{km s$^{-1}$ kpc$^{-2}$},\\
 \Omega^{''}_0 =~0.671\pm0.011~\hbox{km s$^{-1}$ kpc$^{-3}$}.
 \end{array}
 \end{equation}
Here, $V_0=240.6\pm3.0$~km s$^{-1}$ for the adopted $R_0=8.1\pm0.1$~kpc. 

In the second step, we form the velocities $V_R$ and $\Delta V_{circ}$ using the parameters found in the solution (\ref{sol-I}). Next, we perform a spectral analysis of the velocities $V_R$ and $\Delta V_{circ}$. The results are shown in Fig.~\ref{f-1725-RT},
where the velocities $V_R$ and $\Delta V_{circ}$ as a function of distance $R$ for the sample of OB2 stars with line-of-sight velocities and the corresponding power spectra are presented. The following estimates were obtained from these data: $f_R=4.8\pm0.7$~km s$^{-1}$, $f_\theta=4.1\pm0.9$~km s$^{-1}$, $\lambda_R=2.1\pm0.2$~kpc ($i_R=-9.4\pm0.9^\circ$ for $m=4$), $\lambda_\theta=2.2\pm0.4$~kpc ($i_\theta=-9.8\pm1.8^\circ$ for $m=4$), $(\chi_\odot)_R=-116\pm12^\circ$, and $(\chi_\odot)_\theta=-156\pm14^\circ$.

In the third step, we use the entire sample of 9750 OB2 stars with proper motions and parallaxes. Given
relations (\ref{DelVRot}) and (\ref{a-04}) and that $\lambda$ and $\chi_\odot$ are already known, Eq. (\ref{EQ-2}) can be written in a more convenient form for the direct determination of $f_R$ and $f_\theta$:
 \begin{equation}\begin{array}{lll}
 V_l= U_\odot\sin l-V_\odot\cos l
 -r\Omega_0\cos b\\
 +(R-R_0)(R_0\cos l-r\cos b)\Omega^\prime_0\\
 +0.5(R-R_0)^2(R_0\cos l-r\cos b)\Omega^{\prime\prime}_0,\\
 -f_R \cos\chi     \sin(l+\theta)
 +f_\theta\sin\chi \cos(l+\theta),
 \label{EQ-22}
 \end{array}\end{equation}
where
 $$ 
 \chi= {2\pi R_0\over\lambda}\ln(R/R_0)-m\theta+\chi_\odot.
 \label{chi-2}
 $$ 
As a result of the LSM solution of the conditional equations (\ref{EQ-22}), we found the two velocities $(U,V)_\odot=(6.18,6.40)\pm(0.25,0.51)$~km s$^{-1}$ and
 \begin{equation}
 \label{sol-II}
 \begin{array}{lll}
      \Omega_0 =30.98\pm0.17~\hbox{km s$^{-1}$ kpc$^{-1}$},\\
  \Omega^{'}_0 =-4.175\pm0.027~\hbox{km s$^{-1}$ kpc$^{-2}$},\\
 \Omega^{''}_0 =~0.697\pm0.011~\hbox{km s$^{-1}$ kpc$^{-3}$},\\
       f_R = 4.43\pm0.56~\hbox{km s$^{-1}$},\\
  f_\theta = 1.30\pm0.62~\hbox{km s$^{-1}$}
 \end{array}
 \end{equation}
with the adopted $\lambda=2.1$~kpc and $\chi_\odot=-120^\circ$. In this solution 
$V_0=250.9\pm3.4$~km s$^{-1}$ for $R_0=8.1\pm0.1$~kpc.
 
Note that $f_R$ and $f_\theta$ depend very strongly on the adopted phase of the Sun $\chi_\odot$. The effect is illustrated
by the data in Table.~\ref{t:01}, where the parameters of the model (\ref{EQ-22}) obtained at four values of $\chi_\odot$ are given. The values of $f_R$ and $f_\theta$ found are clearly seen to be related to the velocities $U_\odot$ and $V_\odot$. This is because $\sin l$ and $\sin(l+\theta)$ as well as $\cos l$ and $\cos(l+\theta)$ at small angles $\theta$ are close (see Eq. (\ref{EQ-22})). Such a relation between these parameters was studied in detail in Bobylev and Bajkova (2014b).

We will also give the solution obtained as a result of the LSM solution of the conditional equations (\ref{EQ-22}) with the adopted $\lambda=2.1$~kpc and $\chi_\odot=-125^\circ$. In it we found the two velocities $(U,V)_\odot=(5.69,7.28)\pm(0.26,0.46)$~km s$^{-1}$ and
 \begin{equation}
 \label{sol-III}
 \begin{array}{lll}
      \Omega_0 =31.03\pm0.16~\hbox{km s$^{-1}$ kpc$^{-1}$},\\
  \Omega^{'}_0 =-4.183\pm0.027~\hbox{km s$^{-1}$ kpc$^{-2}$},\\
 \Omega^{''}_0 =~0.699\pm0.011~\hbox{km s$^{-1}$ kpc$^{-3}$},\\
       f_R = 5.07\pm0.55~\hbox{km s$^{-1}$},\\
  f_\theta = 0.11\pm0.58~\hbox{km s$^{-1}$}.
 \end{array}
 \end{equation}
We chose the solutions (\ref{sol-II}) and (\ref{sol-III}) as the best ones for the following reasons. First, the most realistic phases of the Sun $\chi_\odot=-120^\circ$ and $\chi_\odot=-125^\circ$ for this sample of stars were used in these solutions. Indeed, based on a spectral analysis, this angle is determined more accurately from the radial velocities, $(\chi_\odot)_R=-116^\circ$. A value that satisfies both radial and tangential velocity components, i.e., close to the mean $[(\chi_\odot)_R+(\chi_\odot)_\theta]/2=-136^\circ$, must enter into Eq. (\ref{EQ-22}). For the spiral pattern shown in Fig.~\ref{f-1812-XY} the Sun’s phase is known accurately, $\chi_\odot=-140^\circ$ (Bobylev and Bajkova 2014a). Thus, $\chi_\odot$ should be chosen from the range $[-116^\circ,-136^\circ]$.

Second, it is important to have the correct relation between the velocities $U_\odot$ and $V_\odot$. We are oriented to those obtained in the solution (\ref{sol-I}). As can be seen from Table~\ref{t:01}, the velocities $U_\odot$ and $V_\odot$ can change greatly with the adopted phase $\chi_\odot$. The main conclusion from our analysis of the table is that in this method we can obtain velocity perturbations very far from reality when using an erroneous phase angle of the Sun. Note that the parameters derived with the phase $\chi_\odot=-130^\circ$ still meet our requirements.

An analysis of the results of our solution of the conditional equations (\ref{EQ-22}) leads to a number of conclusions. The most probable phase of the Sun $\chi_\odot$ for the sample of OB2 stars under consideration lies in the range $[-120^\circ,-130^\circ]$. The amplitude of the radial perturbations $f_R$ lies in the range $[4.4,5.9]$~km s$^{-1}$ and is determined with errors of $\pm0.6$~km~s$^{-1}$.

As regards the angular velocity of Galactic rotation $\Omega_0$ and its derivatives $\Omega^{\prime}_0$ and $\Omega^{\prime\prime}_0$, they are well determined both without the parameters $f_R$ and $f_\theta$ and with their inclusion in the list of unknowns being determined. In comparison with the solution (\ref{sol-I}), in the solution (\ref{sol-II}) and Table~\ref{t:01} the error in $\Omega_0$ increased significantly.

 \begin{table}[t] \caption[]{\small
The kinematic parameters found from OB2 stars based on Eq. (\ref{EQ-22}). 
 }
  \begin{center}  \label{t:01}
  \small
  \begin{tabular}{|l|r|r|r|r|r|}\hline
  Parameters  &
  $\chi_\odot=-100^\circ$ & $\chi_\odot=-110^\circ$ & $\chi_\odot=-130^\circ$ & $\chi_\odot=-140^\circ$\\\hline

   $U_\odot,$ km s$^{-1}$ &  $ 7.79\pm0.18$ & $ 7.15\pm0.22$ & $  5.07\pm0.27$& $  4.30\pm0.29$\\
   $V_\odot,$ km s$^{-1}$ &  $-0.34\pm0.68$ & $ 3.61\pm0.60$ & $  8.02\pm0.42$& $  8.61\pm0.35$\\
  $\Omega_0,$
     km s$^{-1}$ kpc$^{-1}$      &  $30.02\pm0.17$ & $30.65\pm0.17$ & $ 31.22\pm0.15$& $ 31.18\pm0.14$\\
  $\Omega^{'}_0,$
    km s$^{-1}$ kpc$^{-2}$ & $-4.089\pm0.027$&$-4.149\pm0.028$&$-4.198\pm0.027$& $-4.193\pm0.025$\\
 $\Omega^{''}_0,$
    km s$^{-1}$ kpc$^{-3}$ & $ 0.669\pm0.011$& $0.688\pm0.011$& $0.703\pm0.011$& $ 0.701\pm0.011$\\

    $f_R,$    km s$^{-1}$ &  $-0.39\pm0.57$ & $ 2.55\pm0.58$ & $  5.86\pm0.54$& $  6.32\pm0.51$\\
  $f_\theta,$ km s$^{-1}$ &  $ 9.13\pm0.76$ & $ 4.69\pm0.70$ & $ -1.03\pm0.55$& $ -2.36\pm0.49$\\

 \hline
 \end{tabular}\end{center} \end{table}

 \section*{DISCUSSION}
 \subsection*{Galactic Rotation Parameters}
The angular velocity of Galactic rotation $\Omega_0$ and its derivatives $\Omega^{\prime}_0$ and $\Omega^{\prime\prime}_0$ found in the solution (\ref{sol-I}) are typical for very young objects in the Galactic thin disk and are in excellent agreement with their estimates obtained by other authors.
 
For example, based on 130 Galactic masers with measured trigonometric parallaxes, Rastorguev
et al. (2017) found the solar velocity components $(U_\odot,V_\odot)=(11.40,17.23)\pm(1.33,1.09)$~km s$^{-1}$ and the following parameters of the Galactic rotation curve: $\Omega_0=28.93\pm0.53$~km s$^{-1}$ kpc$^{-1}$ , $\Omega^{'}_0=-3.96\pm0.07$~km s$^{-1}$ kpc$^{-2}$ , and $\Omega^{''}_0=0.87\pm0.03$~km s$^{-1}$ kpc$^{-3}$ , where $V_0=243\pm10$~km s$^{-1}$ for $R_0=8.40\pm0.12$~kpc found.

Based on a sample of 147 masers, Reid et al. (2019) found the following values of the two most important kinematic parameters: $R_0=8.15\pm0.15$~kpc and $\Omega_\odot=30.32\pm0.27$~km s$^{-1}$ kpc$^{-1}$, where $\Omega_\odot=\Omega_0+V_\odot/R$. The velocity $V_\odot=12.2$~km s$^{-1}$ was taken from Sch\"onrich et al. (2010). These authors used an expansion of the linear Galactic rotation velocity into a series.

Based on the proper motions of $\sim$6000 OB stars from the list by Xu et al. (2018) with proper motions and parallaxes from the Gaia DR2 catalogue, Bobylev and Bajkova (2019) found $(U_\odot,V_\odot)=(6.53,7.27)\pm(0.24,0.31)$~km s$^{-1}$, $\Omega_0 =29.70\pm0.11$~km s$^{-1}$ kpc$^{-1}$, $\Omega^{'}_0=-4.035\pm0.031$~km s$^{-1}$ kpc$^{-2}$, and $\Omega^{''}_0=0.620\pm0.014$~km s$^{-1}$ kpc$^{-3}$, where $V_0=238\pm5$~km s$^{-1}$ for the adopted $R_0=8.0\pm0.15$~kpc.

 \begin{table}[t] \caption[]{\small
The parameters of the Galactic spiral density wave estimated by various authors. 
 }
  \begin{center}  \label{t:02}
  \small
  \begin{tabular}{|l|c|c|c|c|c|c|c|}\hline
 Sample & Ref & $f_R,$ km s$^{-1}$ & $f_\theta,$ km s$^{-1}$ & $\lambda,$ kpc & $i,$ deg.
   & $\chi_\odot,$ deg. & $m$ \\\hline

 OB stars, Cep., OSCs & [1]& $3.6\pm0.4$ &$4.7\pm0.6$ & $$   &$- 4.2\pm0.2$ &$-165\pm1$& 2\\
 Cepheids     & [2]& $6.3\pm2.4$ &$4.4\pm2.4$ &   $$      &$-6.8\pm0.7$  &$-70\pm16$& 2\\
 Cepheids     & [3]& $3.5\pm1.7$ &$7.5\pm1.8$ &   $$      &$-11.4\pm12$  &$-20\pm9$ & 4\\
 OB associations& [4]& $6.6\pm1.4$ &$1.8\pm1.4$ &$2.0\pm0.2$&  $$          &$$        &  \\
 Cepheids     & [5]& $6.7\pm2.3$ &$1.4\pm1.6$ &   $$      &$- 6.0\pm0.7$ &$-85\pm15$& 2\\
 OSCs          & [5]& $5.5\pm2.3$ &$0.2\pm1.6$ &   $$      &$-12.2\pm0.7$ &$-88\pm15$& 4\\
 OB stars     & [5]& $6.6\pm2.5$ &$0.4\pm2.3$ &   $$      &$- 6.6\pm0.9$ &$-97\pm18$& 2\\
 OSCs, HI, HII & [6]& $5.9\pm1.1$ &$4.6\pm0.5$ &$2.1\pm0.5$&  $$ & $-119~~~~~~$ & \\
 Masers       & [7]& $7.7\pm1.6$ &  $$ &   $2.2\pm0.3$   &$-5.0\pm0.5$  &$-147\pm10$& 2\\
 Masers       & [8]& $6.9\pm1.4$ &$2.8\pm1.0$ &          &$-10.4\pm0.3$ &$-125\pm10$& 4\\
 OB stars    & [9]& $7.1\pm0.3$ &$6.5\pm0.4$ &$2.8\pm0.2$ &$$ &$-128\pm6$&  4\\
 OSCs          &[10]& $4.6\pm0.7$ &$1.1\pm0.4$ &   $$      & $$ &  $$  & 4\\
 \hline
 This paper I  &&$4.8\pm0.7$ &$4.1\pm0.9$ &$2.1\pm0.2$ &$-9.4\pm0.9$ &$-116\pm12$& 4\\
 This paper II &&$4.4\pm0.6$ &$1.3\pm0.6$ &$2.1~~~~~~$ &$$ &$[-120,-130]$ & 4\\
 \hline
 \end{tabular}\end{center}
  {\small
 [1]Byl and Ovenden (1978);
 [2]Mishurov et al. (1997);
 [3]Mishurov and Zenina (1999);
 [4]Mel’nik et al. (2001); [
 [5]Zabolotskikh et al. (2002);
 [6]Bobylev et al. (2008);
 [7]Bajkova and Bobylev (2012);
 [8]Rastorguev et al. (2017);
 [9]Bobylev et al. (2018);
 [10]Loktin and Popova (2019).
  }
\end{table}

 \subsection*{Density Wave Parameters}
Table~\ref{t:02} gives the Galactic spiral density wave parameters found by various authors using various observational data. 

Byl and Ovenden (1978) determined these parameters based on a sample of 797 stars of spectral types
from O7 to A5; they also invoked 145 Cepheids and 76 OSCs. A total of 1018 line-of-sight velocities of
these young objects were used.

Mishurov et al. (1997) and Mishurov and Zenina (1999) considered a sample of about 120 Cepheids.
Both their line-of-sight velocities and proper motions were used. Here, the minimum of the $\chi^2$ was searched for to determine the velocity perturbations $f_R$ and $f_\theta$, while simultaneously two variables, $i$ and $\chi_\odot$, were varied. In the original papers the Sun’s phase was measured from the Perseus arm, but we brought these value to our measurement method.

Mel’nik et al. (2001) analyzed the kinematics of a sample of 70 OB associations within 3 kpc of the
Sun. Both line-of-sight velocities and proper motions of these associations were used.

Zabolotskikh et al. (2002) considered various young Galactic objects. Their kinematic sample included
113 classical Cepheids with pulsation periods longer than 9 days, 89 young ($\log t<7.6$) OSCs, 102
blue supergiants as well as the line-of-sight velocities of HI clouds at tangential points and the line-of-sight velocities of HII regions.

Bobylev et al. (2008) used data on young ($\leq$50 Myr) OSCs, the line-of-sight velocities of HI clouds at tangential points, and the line-of-sight velocities of HII regions. As a result, they obtained fairly good coverage of the inner Galaxy. A Fourier analysis slightly differing from the one used in this paper was applied.

Bajkova and Bobylev (2012) used a sample of 44 Galactic masers with measured trigonometric parallaxes.
Rastorguev et al. (2017) already considered a sample of 131 masers. The kinematic parameters of
the spiral structure were determined in the same way as in this paper, based on a spectral analysis. As in this paper, Bobylev et al. (2018) considered a sample of 495 OB stars with data from the Gaia DR2 catalogue, where they searched for the parameters separately from the radial and tangential velocity components based on a spectral analysis. Therefore, the average values of $\lambda$ and $\chi_\odot$ are given in the table.

Loktin and Popova (2019) analyzed the kinematics of $\sim$1000 OSCs from the “Homogeneous Catalog
of Open Cluster Parameters” (Loktin and Popova 2019) with stellar proper motions from the Gaia DR2
catalogue. The line-of-sight velocities were measured for 522 OSCs from this sample.

The lower two rows in Table~\ref{t:02} give the parameters that in this paper (a) were determined from the stellar radial velocities based on a spectral analysis, designated as method I, and (b) were derived based on the solution of Eq. (\ref{EQ-22}), designated as method II. On the whole, it can be seen that there is good agreement of $f_R$ and $\lambda$ and the angles $i$ and $\chi_\odot$ found in this paper with their other determinations. There is poor agreement only in determining $f_\theta$. However, it can be seen from the table that this parameter is determined unreliably by any methods.

\section*{CONCLUSIONS}
To study the kinematics of the Galaxy, we used the sample of OB2 stars from Xu et al. (2021) with
proper motions and trigonometric parallaxes from the Gaia EDR3 catalogue. For 1812 stars from this
sample there are line-of-sight velocities taken from published sources.

Based on the proper motions and parallaxes for the entire sample of 9750 OB2 stars, we found the velocities $(U,V)_\odot=(7.17,7.37)\pm(0.16,0.24)$~km s$^{-1}$ and the components of the angular velocity of Galactic rotation: $\Omega_0 =29.700\pm0.076$~km s$^{-1}$ kpc$^{-1}$, $\Omega^{'}_0 =-4.008\pm0.022$~km s$^{-1}$ kpc$^{-2}$, and $\Omega^{''}_0 =~0.671\pm0.011$~km s$^{-1}$ kpc$^{-3}$, where the linear rotation velocity of the Galaxy at the solar distance is $V_0=240.6\pm3.0$~km s$^{-1}$ for the adopted $R_0=8.1\pm0.1$~kpc. The values of these parameters are typical for young objects in the Galactic thin disk and are in excellent agreement with their estimates obtained by other authors. However, in our case, owing to the use of a huge number of stars, they were determined with very small errors.

Based on 1812 OB2 stars with line-of-sight velocities, we calculated the space velocities $V_R$
and $\Delta V_{circ}$. We performed a spectral analysis independently for the radial and residual tangential velocities. The following estimates were obtained: $f_R=4.8\pm0.7$~km s$^{-1}$, $f_\theta=4.1\pm0.9$~km s$^{-1}$, $\lambda_R=2.1\pm0.2$~kpc ($i_R=-9.4\pm0.9^\circ$ for $m=4$), $\lambda_R=2.1\pm0.2$~kpc ($i_\theta=-9.8\pm1.8^\circ$ for $m=4$), $(\chi_\odot)_R=-116\pm12^\circ$, and $(\chi_\odot)_\theta=-156\pm14^\circ$. We see that these parameters are determined more reliably from
the stellar radial velocities.

The basic kinematic equation was also solved by including the velocity perturbations $f_R$ and $f_\theta$ as additional unknowns. As a result, we concluded that the most probable phase of the Sun $\chi_\odot$ for the sample of OB2 stars under consideration lies in the range $[-120^\circ,-130^\circ]$. The radial velocity perturbation $f_R$ lies in the range 
$[4.4,5.9]$~km s$^{-1}$ and is determined by this method with errors of $\pm0.6$~km s$^{-1}$.

On the whole, we concluded that there is good agreement of the parameters $f_R$ and $\lambda$ and the angles $i$ and $\chi_\odot$ found in this paper by two methods both between themselves and with their determinations by other authors. There is poorer agreement in determining $f_\theta$. We have more trust in the approach using a spectral analysis, $f_\theta=4.1\pm0.9$~km s$^{-1}$. Thus, both velocity perturbation amplitudes are nonzero at a high significance level.

Note that the method of estimating the velocity perturbations $f_R$ and $f_\theta$ based on the kinematic model (\ref{EQ-22}) was proposed by Cr\'ez\'e and Mennessier (1973) with a number of simplifications that limit the range of application of the method. In particular, the ratio $R/R_0$ was expanded into a series using only the first term of the series.

The method based on a periodogram Fourier analysis (Bajkova and Bobylev 2012) takes into account
both the logarithmic behavior of the Galactic spiral structure and the positions angles of the objects. It does not use any simplifications and assumptions, which allows the velocities of the objects distributed in a wide range of Galactocentric distances to be analyzed most accurately. Thus, this method is more reliable.

\section*{ACKNOWLEDGMENTS}
We are grateful to Yu.N. Mishurov for the useful and in-depth discussion of our results.

\newpage
 \bigskip\medskip{\bf\Large REFERENCES}\medskip{\small

 \begin{enumerate}

 \item
 R. Abuter, A. Amorim, N. Baub\"ock, J. P. Berger,
H. Bonnet, W. Brandner, Y. Cl\'enet, V. Coud\'e du
Foresto, et al. (GRAVITY Collab.), Astron. Astrophys.
{\bf 625}, L10 (2019).

 \item
A. T. Bajkova and V. V. Bobylev, Astron. Lett. {\bf 38}, 549
(2012). 

 \item
D. A. Barros, A. Perez-Villegas, T. A. Michtchenko,
and J. R. D. L\'epine, Front. Astron. Space Sci. {\bf 8}, 48
(2021). 

 \item
V. V. Bobylev, A. T. Bajkova, and A. S. Stepanishchev,
Astron. Lett. {\bf 34}, 515 (2008)].

 \item
V. V. Bobylev and A. T. Bajkova, Mon. Not. R. Astron.
Soc. {\bf 437}, 1549 (2014a). 

 \item
V. V. Bobylev and A. T. Bajkova, Mon. Not. R. Astron.
Soc. {\bf 441}, 142 (2014b). 

\item
V. V. Bobylev and A. T. Bajkova, Astron. Lett. {\bf 44}, 676
(2018).

 \item
V. V. Bobylev and A. T. Bajkova, Astron. Lett. {\bf 45}, 331
(2019). 

 \item
V. V. Bobylev and A. T. Bajkova, Astron. Rep. {\bf 65}, 498
(2021). 

 \item
A. G. A. Brown, A. Vallenari, T. Prusti, J. H. J. de
Bruijne, C. Babusiaux, C. A. L. Bailer-Jones,
M. Biermann, D. W. Evans, et al. (Gaia Collab.),
Astron. Astrophys. {\bf 616}, 1 (2018).

 \item
A. G. A. Brown, A. Vallenari, T. Prusti, J. H. J. de
Bruijne, C. Babusiaux, M. Biermann, O.L. Creevely,
D.W. Evans, et al. (Gaia Collab.), Astron. Astrophys.
{\bf 649}, 1 (2021). 

 \item
W. B. Burton, Astron. Astrophys. {\bf 10}, 76 (1971). 

 \item
J. Byl and M. W. Ovenden, Astrophys. J. {\bf 225}, 496
(1978). 

 \item
M. Cr\'ez\'e and M. O. Mennessier, Astron. Astrophys.
{\bf 27}, 281 (1973). 

 \item
D. Fern\'andez, F. Figueras, and J. Torra, Astron. Astrophys.
{\bf 372}, 833 (2001). 

 \item
E. Griv, C.-C. Lin, C.-C. Ngeow, and I.-G. Jiang,
New Astron. {\bf 29}, 9 (2014). 

 \item
E. Griv and I.-G. Jiang, Astron. Nachr. {\bf 336}, 196
(2015). 

 \item
 J. R. D. L\'epine, Yu. Mishurov and S. Yu. Dedikov,
Astrophys. J. {\bf 546}, 234 (2001).

 \item
C. C. Lin and F. H. Shu, Astrophys. J. {\bf 140}, 646
(1964). 

 \item
C. C. Lin, C. Yuan, and F. H. Shu, Astrophys. J. {\bf 155},
721 (1969). 

 \item
A. V. Loktin and M. E. Popova, Astrophys. Bull. {\bf 72},
257 (2017). 

 \item
A. V. Loktin and M. E. Popova, Astrophys. Bull. {\bf 74},
270 (2019). 

 \item
A. M. Mel’nik, A. K. Dambis, and A. S. Rastorguev,
Astron. Lett. {\bf 27}, 521 (2001). 

 \item
Yu. N. Mishurov, I. A. Zenina, A. K. Dambis,
A. M. Mel’nik, and A. S. Rastorguev, Astron. Astrophys.
{\bf 323}, 775 (1997). 

 \item
Yu. N.Mishurov and I. A. Zenina, Astron. Astrophys.
{\bf 341}, 81 (1999). 

 \item
M. E. Popova and A. V. Loktin, Astron. Lett. {\bf 31}, 663
(2005). 

\item
T. Prusti, J.H. J. de Bruijne,A.G. A. Brown, A. Vallenari,
C. Babusiaux, C. A. L. Bailer-Jones, U. Bastian,
M. Biermann, et al. (Gaia Collab.), Astron. Astrophys.
{\bf 595}, 1 (2016).

\item
A. S. Rastorguev, M. V. Zabolotskikh, A. K. Dambis,
N. D. Utkin, V. V. Bobylev, A. T. Bajkova, Astrophys.
Bull. {\bf 72}, 122 (2017).

 \item
M. J. Reid, K. M. Menten, A. Brunthaler,
X. W. Zheng, T. M. Dame, Y. Xu, J. Li, N. Sakai,
et al., Astrophys. J. {\bf 885}, 131 (2019). 

 \item
K. Rohlfs, Lectures on Density Wave Theory
(Springer, Berlin, 1977). 

 \item
D. Russeil, Astron. Astrophys. {\bf 397}, 133 (2003). 

 \item
R. Sch\"onrich, J. J. Binney, andW. Dehnen,Mon.Not.
R. Astron. Soc. {\bf 403}, 1829 (2010). 

 \item
A. Siebert, B. Famaey, J. Binney, B. Burnett,
C. Faure, I. Minchev, M. E. K. Williams,  et al., Mon. Not. R. Astron. Soc. {\bf 425}, 2335
(2012). 

 \item
B. A. Skiff, VizieR Online Data Catalog, B/mk
(2014). 

 \item
Y. Xu, S. B. Bian, M. J. Reid, J. J. Li, B. Zhang,
Q. Z. Yan, T.M.Dame, K. M. Menten, et al., Astron.
Astrophys. {\bf 616}, L15 (2018). 

 \item
Y. Xu, L.G.Hou, S. Bian, C. J.Hao, D. J. Liu, J. J. Li,
and Y. J. Li,Astron. Astrophys. {\bf 645}, L8 (2021). 

 \item
M. V. Zabolotskikh, A. S. Rastorguev, and
A. K. Dambis, Astron. Lett. {\bf 28}, 454 (2002). 

 \end{enumerate} }
\end{document}